\documentclass[showpacs,aps,prb,twocolumn,superscriptaddress]{revtex4-1}
\usepackage{graphicx} 
\usepackage{dcolumn}
\usepackage{bm}
\usepackage{amssymb,amsmath}
\usepackage{epstopdf}
\usepackage{xcolor} 
\usepackage[normalem]{ulem} 
\begin{document}
\title{Guided Mode Resonances in Flexible 2D THz Photonic Crystals}
\normalsize
\author{C. Kyaw}
\affiliation{Department of Physics and Astronomy, Howard University, Washington, DC 20059, USA}
\author{R.~Yahiaoui}
\affiliation{Department of Physics and Astronomy, Howard University, Washington, DC 20059, USA}
\author{Z. A.~Chase}
\affiliation{Department of Physics and Astronomy, Howard University, Washington, DC 20059, USA}
\author{V.~Tran}
\affiliation{Department of Physics and Astronomy, Howard University, Washington, DC 20059, USA}
\author{A.~Baydin}
\affiliation{Department of Electrical and Computer Engineering, Rice University, Houston, TX 70005, USA}
\author{F.~Tay}
\affiliation{Department of Electrical and Computer Engineering, Rice University, Houston, TX 70005, USA}
\author{J.~Kono}
\affiliation{Department of Electrical and Computer Engineering, Rice University, Houston, TX 70005, USA}
\affiliation{Department of Physics and Astronomy, Rice University, Houston, TX 70005, USA}
\affiliation{Department of Materials Science and Nanoengineering, Rice University, Houston, TX 70005, USA}
\author{M.~Manjappa}
\affiliation{Division of Physics and Applied Physics, School of Physical and Mathematical Sciences, Nanyang Technological University, 21 Nanyang Link, Singapore 637371, Singapore}
\author{R.~Singh}
\affiliation{Division of Physics and Applied Physics, School of Physical and Mathematical Sciences, Nanyang Technological University, 21 Nanyang Link, Singapore 637371, Singapore}
\author{D. C.~Abeysinghe}
\affiliation{Air Force Research Laboratory, Materials and Manufacturing Directorate, Wright-Patterson Air Force Base, OH 45433, USA}
\author{A. M.~Urbas}
\affiliation{Air Force Research Laboratory, Materials and Manufacturing Directorate, Wright-Patterson Air Force Base, OH 45433, USA}
\author{T. A.~Searles}
\email[]{thomas.searles@howard.edu}
\thanks{corresponding author.}
\affiliation{Department of Physics and Astronomy, Howard University, Washington, DC 20059, USA}
\date{\today}

\begin{abstract}
An article usually includes an abstract, a concise summary of the work
covered at length in the main body of the article. 
\begin{description}
\item[Usage]
Secondary publications and information retrieval purposes.
\item[Structure]
You may use the \texttt{description} environment to structure your abstract;
use the optional argument of the \verb+\item+ command to give the category of each item. 
\end{description}
\end{abstract}
\begin{abstract}
In terahertz (THz) photonics, there is an ongoing effort to develop thin, compact devices such as dielectric photonic crystal (PhC) slabs with desirable light matter interactions. However, previous works in THz PhC slabs are limited to rigid substrates with thicknesses $\sim$ 100s of micrometers. Dielectric PhC slabs have been shown to possess in-plane modes that are excited by external radiation to produce sharp guided mode resonances with minimal absorption for applications in sensors, optics and lasers. 
Here, we confirm the existence of guided resonances in a membrane-type THz PhC slab with subwavelength ($\lambda_{0}$/6 - $\lambda_{0}$/12) thicknesses of flexible dielectric polyimide films. The transmittance of the guided resonances was measured for different structural parameters of the unit cell. Furthermore, we exploited the flexibility of the samples to  modulate the linewidth of the guided modes down to 1.5 GHz for bend angle of $\theta$ $\geq 5^{\circ}$; confirmed experimentally by the suppression of these modes. The mechanical flexibility of the device allows for an additional degree of freedom in system design for optical components for high-speed communications, soft wearable photonics and implantable medical devices. 
\end{abstract}

\maketitle

\section{Introduction\vspace{-1em}}
Communications specific applications such as Wi-Fi enabled video calls, large data transfers and entertainment streaming have pushed the bandwidth limits of non-optical components like copper wire, cable connections, etc.  To achieve  higher bandwidths ($\geq$ 100 Gbit/s) required by the increased and future demand, researchers aim to develop new technologies designed to operate at terahertz (THz) frequencies. However, the majority of current devices and components are large and bulky with slow improvement for practical, compact THz platforms. 
The most popular candidates for planar, compact THz light manipulation are plasmonic metamaterials; sub-wavelength arrays of metallic resonators periodically distributed on a dielectric substrate. Applications of planar metamaterials include  nonlinear enhancements \cite{Merbold:11, Bagiante2015GiantEF,Kim18}, near-field focusing \cite{Wang15flatlens,Guven2006} and beam steering devices \cite{Reis15,Liu18}. Additionally, due to the use of polyimide \cite{CongPolyimide2014, Burrow17, YahiaouiPRB18}, PDMS \cite{Li2013} and carbon nanotubes \cite{Hong2013} substrates, THz metasurfaces are desirable for flexible photonics applications. But, they are limited by intrinsic Ohmic loss of metals at THz frequencies and requirements for fabrication of multi-layer or composite structures.\\ 
\indent To avoid the above limitations, photonic crystal (PhC) slabs have been employed to manipulate light-matter interaction with THz waves. In general, PhC slabs are two-dimensional (2D) planar devices composed of a dielectric material with an array of holes, and periodicity in the range of the wavelength of interest. These planar 2D PhC slabs exhibit strong field confinement and interaction with THz waves with minimal absorption loss.  In addition, the dielectric platforms of PhC slabs offer low-loss alternatives to metallic resonators.\\ 
\indent As such, PhC slabs have been used to guide \cite{Lin1998,Mekis1996}, filter \cite{Blanco2000,Krauss1996} and enhance electric field distributions in small volumes \cite{Lalouat2011,Mjumdar2007}. Although, light is usually confined within PhC slabs, certain guided modes can strongly couple with external radiation when they are normally incident to an in-plane resonance on the PhC surface. 
The in-plane mode, also described as photonic band edge effect or distributed feedback effect, can be utilized for lasing in photonic crystals\cite{Noda2001,Meier}. Experimental demonstrations of the properties of 2D THz PhC slabs have been reported in silicon \cite{Prasad:08,Prasad07, Daniel05, Daniel06, DanielJian06, Kakimi2014, Tsuruda:15, Sherwin2003,Suminokura2014,Otter2014,Yata2016,Yee07,Yee09} and GaAs \cite{Sherwin06} media.  The majority of reported THz PhC slabs have  thicknesses in the range of THz wavelength \cite{Prasad:08,Prasad07, Daniel05, Daniel06, DanielJian06, Kakimi2014, Tsuruda:15, Sherwin2003,Suminokura2014,Otter2014,Yata2016, Sherwin06} with few demonstrations around 50 micron thickness \cite{Yee07,Yee09}. For thin samples originating from rigid crystals, the PhC slabs are fragile and difficult to handle.\\  
\indent In this Letter, we report the existence of guided mode resonances supported by a flexible membrane-type PhC (both 50 and 25 $\mu$m thick) operating at THz frequencies. Experimental measurements of the fabricated device confirm the existence of guided resonances that couple the in-plane resonance of the PhC slab with external radiation. Systematic studies of these resonances with respect to different thicknesses and hole geometries are presented. Curvature dependent spectral measurements show active tunability of the modulation depth and quality factor of the guided modes. Our results pave the way for a new flexible medium for dielectric PhC slabs with sub-wavelength thickness. The mechanical robustness of the devices allows for applications in flexible photonics including integration with future biomedical and communications platforms.

\section{Materials and Methods}

\begin{figure}[b]
\centering
\includegraphics[width=\linewidth]{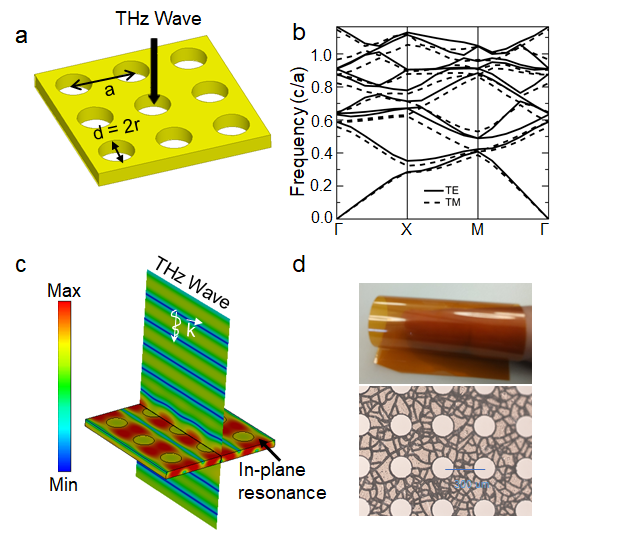}
\caption{Flexible 2D THz Photonic Crystal Slab. (a) Pictorial of the structural parameters of a THz wave normally incident on 2D  photonic crystal (PhC) slab with airhole diameter d or radius r, and period a in a square lattice. (b) Band dispersion  calculated for the PhC slab in (a) with r/a = 0.3 and $\epsilon$ = 3.4.  The TM mode is dashed and the TE mode is solid. (c) Representation of the mechanism for supporting guided modes where the incident THz wave couples with the in-plane resonant mode of the PhC slab. (d) Optical images of the highly flexible Kapton film (top) and the fabricated PhC slab (botttom). Residues from etching process form a mesh pattern around the holes.}
\end{figure}

\begin{figure*}
\centering
    \includegraphics[width=1\textwidth]{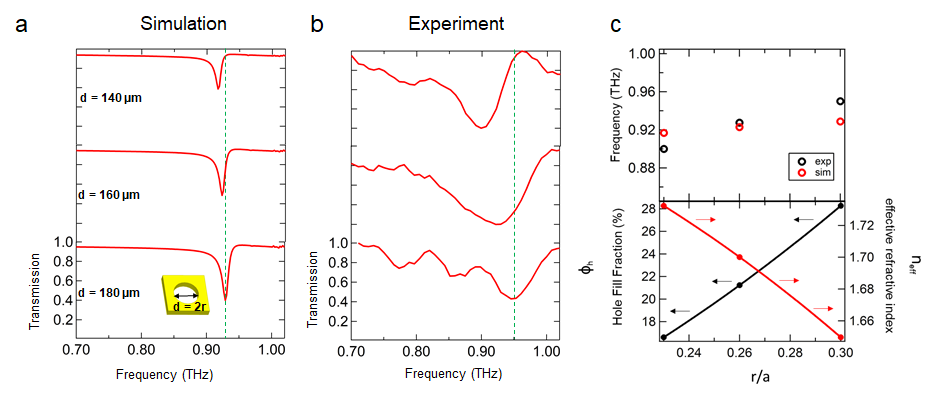}

\caption{\label{fig:wide}Transmission of the guided mode as a function of hole diameter for 50 $\mu$m thick sample. (a) Simulated and (b) measured transmission spectra for different diameters in a square lattice with circular air holes of different diameters in a square lattice. (c) Shifts in frequency  of the guided resonances are extracted from experiment and simulation for increasing r/a values (top). The hole fill fraction (black) and effective refractive index (red) are also calculated for different r/a values (bottom).}
\end{figure*}
A schematic of the primary experimental setup is given in Fig. 1(a) for a THz wave  incident on the PhC slab with air holes of a specific diameter $d$, radius $r$ and period $a$ in a square lattice. Next, we calculated the first ten bands of the photonic band structure using the MIT Photonic Bands (MPB) Package \cite{Johnson:01} for a 2D PhC slab with dielectric constant $\epsilon$ = 3.4 and $r/a$ = 0.3 and plotted for both TE and TM polarizations [Fig. 1(b)]. Unlike silicon ($n_{Si}$ $\sim$ 3.418 at 1 THz) or GaAs ($n_{GaAs}$ $\sim$ 3.59 at 1 Thz) PhC slabs, no clear photonic bandgap is present in our samples made of Kapton polyimide ($n_{K}$ $\sim$ 1.843). However, we observed and verified through spectrographic analysis guided resonances originating from the in-plane resonant mode of the PhC slab coupled to the external THz radiation in the direction normal to the PhC plane [Fig. 1(c)]. The in-plane resonance leaks out from the surface as a guided resonance as it is coupled to the radiation.

Standard polyimide etch recipes reported in literature are not applicable for Kapton films which are uniquely synthesized for high temperature resistivity. To fabricate the air holes in our samples, we first patterned a titanium hard mask on the Kapton film, then dry etched with an inductively coupled plasma (ICP) etcher (Plasma-Pro 100 Cobra). The etching conditions are as follows: oxygen pressure of 20 mTorr, flow rate of 60 sccm, RF1 power at 75 W and RF2 at power 1500 W with an approximate etch rate of $\sim$ 1.05 $\mu$m/min. A photo of a flexible Kapton polyimide film with an optical image of the photonic crystal pattern on the sample is also shown [Fig. 1(d)].

Transmission measurements were performed with standard terahertz time domain spectroscopy (THz-TDS) techniques. Two different experimental setups, at Rice  and at NTU,  were utilized to perform the measurements. The setups used output from a Ti:sapphire oscillator producing nearly (100-150 fs) pulses at 800 nm split into pump and probe beams. The pump and probe optical pulses served as generation and detection of the THz field in ZnTe crystals, respectively. The probe beam was delayed with respect to the pump beam using a delay stage which in turn dictates the resolution of the system. Electric field amplitude and phase of the THz waveform are obtained by scanning the delay stage for a maximum resolution of 40 GHz. All measurements were done inside a box purged with dry air or nitrogen to remove excess water vapor. 

Numerical simulations were performed using a finite element method with periodic boundary conditions to emulate a 2D infinite array of unit cells. To ensure the accuracy of the simulations, the length scale of the mesh was set to $\leq \lambda_{0}$ /10 throughout the simulation domain, where $\lambda_{0}$ (600 $\mu$m) is the central wavelength of the incident radiation. The input and output ports were placed at 3$\lambda_{0}$ from the PhC slab with open boundary conditions. To mimic the experiment for active tunability, curvature dependent simulations were performed by using the appropriate beam deflection angles \cite{Su2019}. In all numerical simulations, the permittivity value of $\epsilon$ = 3.4 was applied for the dielectric Kapton layer.

\section{Results and Discussion}

To study the behavior of the guided mode resonances, we simulated and measured the transmission of a 50 $\mu$m thick Kapton PhC slab with circular air holes of varying diameters [Fig.~2(a)-(b)]. Due to the periodicity of the unit cell  (300 $\mu$m), such structures do not diffract normally incident THz radiation for frequencies less than 1 THz. The single guided resonance below 1 THz is observed between 0.9 and 1 THz in both experiment and simulation for all hole diameters. However, the experimental resonance has a larger linewidth compared to the simulations. The difference between experiment and simulations is due to finite sampling time, scattering losses and possible disorder from slight imperfections or defects in the sample which cause the experimental spectra to be dampened and inhomogeneously broadened.  Recent measurements of comparable sharp resonances in terahertz regime show a similar broadening in experiments for holes array in silicon slabs \cite{Yang2019}. Unlike silicon PhC slabs with thicknesses ranging in hundreds of micrometers \cite{Prasad:08,Prasad07, Daniel05, Daniel06, DanielJian06, Kakimi2014}, Fabry-perot oscillations are not observed in our sample.

In order to quantify the resonances, quality factor (Q-factor) and modulation depth (MD) from the guided modes are extracted. Q-factor is defined as $Q$ = $f_{i}$ /$\Delta$$f_{i}$ where $f_{i}$ is the center frequency and $\Delta$$f_{i}$ is the full width at half maximum (FWHM).
The modulation depth is determined as  $MD = A$ - min $(T)$ where $A$ is the maximum transmission and min $(T)$ is the minimum transmission near the resonance dip.  The average calculated Q-factor and modulation depth of the guided modes from simulations were 83.35 and 0.46 respectively.

To describe the dependence on hole geometry, Fig. 2(c) plots the noticeable frequency shift observed in the simulated and measured spectra with respect to change in the fill fraction of holes ($\phi_{h}$) and effective refractive index of the PhC slab ($n$) as a function of $r/a$. The fill fraction of holes ($\phi_{h}$) is calculated as $(r/a)^2$ while effective refractive index is determined from: \cite{Prasad:08} 
\begin{equation}
\mathit{n_{eff}} = \sqrt{\left(\epsilon_{K} *   \phi_{K}\right)+ \left(\epsilon_{h} *   \phi_{h}\right)}
\end{equation}
where $\epsilon_{K} = 3.4$ for Kapton film and $\epsilon_{h} = 1$ for air holes while $\phi_{K} = 1 - \phi_{h}$.
The calculated resonance shifts towards higher frequencies for about 12 GHz in simulations from r/a = 0.23 (d = 140 $\mu$m) to r/a = 0.3 (d= 180 $\mu$m). This is supported by a larger shift of 50 GHz in the same direction in the experiment. These shifts are accompanied by increased fill fraction of the holes ($\sim$ 12\%) and lowered effective refractive index ($\sim$ 0.08) of the PhC slabs.

\noindent

\begin{figure}[b]
\centering
\includegraphics[width=\linewidth]{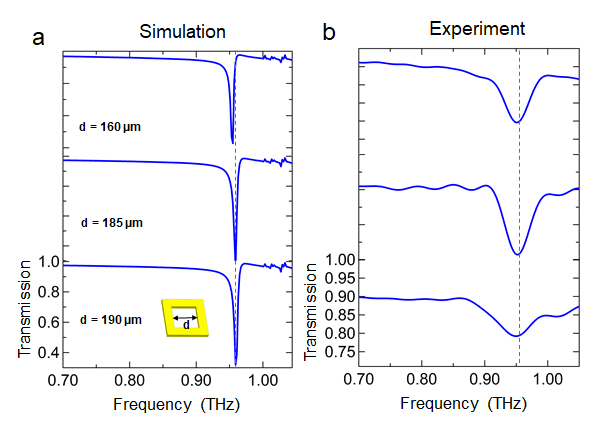}
\caption{Transmission of guided mode as a function of square-hole diameter for 25 $\mu$m sample. (a) Simulated and (b) measured transmission spectra for different diameters in a square lattice.}
\end{figure}

\begin{figure}[!ht]
\centering
\includegraphics[width=\linewidth]{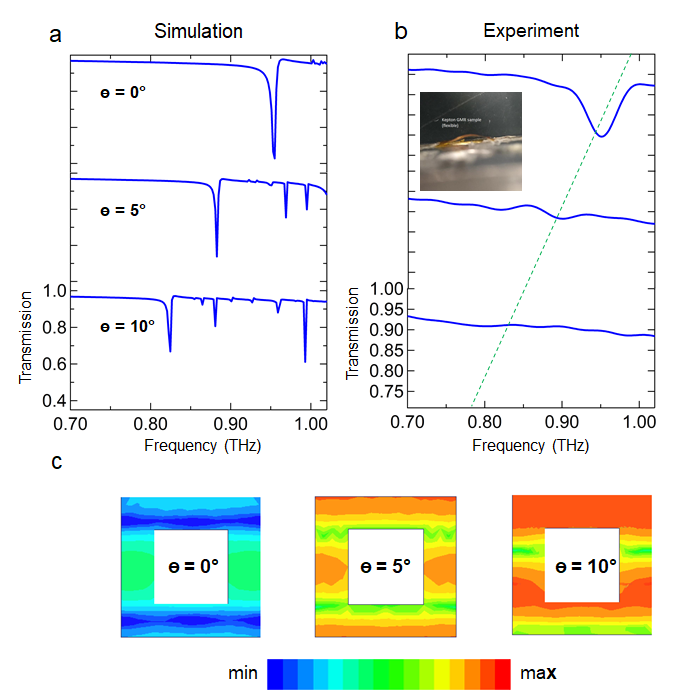}
\caption{Curvature dependent transmission of guided mode. (a) Simulated and (b) measured transmission are plotted for the 25 $\mu$m thick sample with curvature angles $\theta$ = $0^{\circ}$,  $5^{\circ}$, and  $10^{\circ}$. The photo insert describes the setup of curved Kapton sample. The green dashed line highlights the resonance shift. (c) Electric field distributions are mapped at frequencies of the simulated guided resonances for $\theta$ = $0^{\circ}$,  $5^{\circ}$, and  $10^{\circ}$}
\end{figure}

To further analyze the guided modes, we reduced the thickness of the Kapton films to 25 $\mu$m ($\lambda_{0}$/12) and fabricated PhC slabs with square holes in a square lattice. Simulated and measured transmission amplitudes of the sample are plotted for varying hole diameters in Fig. 3(a)-(b) with different scales to highlight the experimental spectra with reduced MD. From simulations, guided resonances that originate from periodic square holes and circular holes have very similar shape and linewidth for the same thickness. However, when the thickness is halved to 25 $\mu$m, the resonances exhibited extremely sharp dips (FWHM linewidth $\sim$ 6 GHz) compared to 50 $\mu$m (FWHM linewidth $\sim$ 14 GHz). The average calculated modulation depth of 0.656 and  Q-factor of 168 are observed for the guided resonances. Experimental resonance dips are resolved until 72\% transmission for d = 185 $\mu$m. Similar to 50 $\mu$m sample, a frequency shift towards lower frequency is observed for decreasing hole diameters but the calculated small shift of 4 GHz is not resolved in experiment.

Next, we take advantage of the flexibility of our samples through the demonstration of  active modulation of guided mode resonance (Fig.~4). The  response of the PhC slabs are simulated for different angles of curvature  [Fig.~4(a)] with their corresponding experimental results; where we taped the 25 $\mu$m sample with d = 160 $\mu$m on a curved surfaces of angles $5^{\circ}$ and $10^{\circ}$ as seen in the photo insert [Fig. 4(b)]. The calculated resonances became sharper with lower MD (0.54 - 0.31) as the angle is increased from $0^{\circ}$ to $10^{\circ}$. The linewidth at 72\% transmission decreases from 6 GHz ($\theta$ = $0^{\circ}$) to 2.5 GHz ($\theta$ = $5^{\circ}$) and 1.5 GHz ($\theta$ = $10^{\circ}$) below the sensitivity of the detection equipment. The resonance also shifts towards lower frequencies by as much as 130 GHz from $\theta$ = $0^{\circ}$ to $\theta$ = $10^{\circ}$.  Previous work on rigid SiNx PhC slabs in the visible and NIR attributed angular dependent shifts to the empty lattice approximation\cite{Crozier2016}.  Similarly, the result of Fig.~4 shows a redshift corresponding to an increase in propagation constant $\beta$ at angles further from normal incidence.   This trend is supported by experimental measurements where the resonance dip becomes very small with a redshift from $\theta$ = $0^{\circ}$ to $\theta$ = $5^{\circ}$ with the complete disappearance at $\theta$ = $10^{\circ}$.

To gain a deeper physical understanding, we also investigated the electric field distributions at each curvature angle of the guided resonance. The field distributions on the unit cell at the resonance frequencies 955 GHz, 883 GHz, and 825 GHz for respective beam angles $\theta$ = $0^{\circ}$, $5^{\circ}$, and $10^{\circ}$ are plotted in Fig.~4(c). Stronger field confinement is observed as expected for sharper resonances as the angle is increased from $\theta$ = $0^{\circ}$ to $\theta$ = $10^{\circ}$. This indicates a weaker coupling to incident wave at higher angles as evidenced by the lower transmission dips. The observed strong sensitivity of the guided modes to the surface curvature of PhC slabs confirms an additional degree of freedom to tune the resonances previously not reported for rigid THz PhC slabs.

\section{Conclusion}

In this paper, we demonstrate the existence of the guided resonances in a membrane-type terahertz PhC slab from a flexible thin dielectric medium of Kapton polyimide films. The resonance originates from the coupling of the in-plane resonant mode on PhC plane to external radiation. The frequency positions of the resonances undergo a redshift with increasing diameters of the air holes. For ultra-thin samples with $\lambda_{0}$ /12 thickness, lower MD and linewidths are observed for increased curvature angles. This allows for tuning of the guided modes with respect to the curvature of the PhC slabs. The temperature resistance and the mechanical robustness of the Kapton films allow for easy integration of these PhC slabs into current microfabrication technology.


\begin{acknowledgments}
Funding for this research comes from the Air Force Office of Scientific Research (FA9550-16-1-0346), the NSF (ECCS-1541959) and the W. K. Keck Foundation. T. A. S. acknowledges support from the CNS Scholars Program and C. K. acknowledges support in the form of the Just-Julian Graduate Research Assistantship.
\end{acknowledgments}

\bibliography{apssamp}

\end{document}